\documentclass[conference]{IEEEtran}
\IEEEoverridecommandlockouts
\usepackage{booktabs} 

\usepackage[english]{babel}
\usepackage{blindtext}
\usepackage{times}
\usepackage{balance} 
\usepackage{amsmath,amsfonts,amssymb}
\usepackage{algorithm}
\usepackage[noend]{algorithmic}
\usepackage{graphicx}
\usepackage{xspace}
\usepackage{multirow}
\usepackage{url}
\usepackage{epsfig,subfigure}
\usepackage{epstopdf}
\usepackage{color}
\usepackage{upgreek}

\newcommand{\squishlisttight}{
 \begin{list}{$\bullet$}
  { \setlength{\itemsep}{0pt}
    \setlength{\parsep}{0pt}
    \setlength{\topsep}{0pt}
    \setlength{\partopsep}{0pt}
    \setlength{\leftmargin}{2em}
    \setlength{\labelwidth}{1.5em}
    \setlength{\labelsep}{0.5em}
} }

\newcommand{\squishnumlist} {
\newcounter{qcounter}
\begin{list}{\arabic{qcounter}.~}{\usecounter{qcounter}} 
{  \setlength{\itemsep}{0pt}
    \setlength{\parsep}{0pt}
    \setlength{\topsep}{0pt}
    \setlength{\partopsep}{0pt}
    \setlength{\leftmargin}{2em}
    \setlength{\labelwidth}{1.5em}
    \setlength{\labelsep}{0.5em}
}}

\newcommand{\squishend}{
  \end{list}
}

\newcommand{\eat}[1]{}

\newcommand{\kw}[1]{{\ensuremath {\mathsf{#1}}}\xspace}

\newcommand{\stitle}[1]{\vspace{1ex} \noindent{\bf #1}}
\long\def\comment#1{}

\newcommand{\black}[1]{\textcolor{black}{#1}}

\newcommand{\xin}[1]{\black{#1}}

\newcommand{\ppdblp}{\kw{PP}-\kw{DBLP}}
\newcommand{\figpath}{.}

\newtheorem{example}{Example}

\begin{document}

\title{PP-DBLP: Modeling and Generating Attributed Public-Private Networks with DBLP}

\author{%
{Xin Huang{\small$^{*}$}, Jiaxin Jiang{\small$^{*}$}, Byron Choi{\small $^{*}$}, Jianliang Xu{\small $^{*}$}, Zhiwei Zhang{\small $^{*}$}, Yunya Song{\small $^{\#}$}}

\vspace{1.6mm}\\
\fontsize{10}{10}\selectfont\itshape
$^{*}$Department of Computer Science, Hong Kong Baptist University \\
$^{\#}$Department of Journalism, Hong Kong Baptist University\\
\fontsize{9}{9}\selectfont\ttfamily\upshape
$^{*}$\{xinhuang, jxjian, bchoi, xujl, cszwzhang\}@comp.hkbu.edu.hk,
$^{\#}$yunyasong@hkbu.edu.hk 
}

\maketitle

\begin{abstract}
In many online social networks (e.g., Facebook, Google+, Twitter, and Instagram), users prefer to hide her/his partial or all relationships, which makes such private relationships not  visible to public users or even friends. This leads to a new graph model called public-private networks, where each user has her/his own perspective of the network including the private connections. 
Recently, public-private network analysis has attracted significant research interest in the literature. A great deal of important graph computing problems (e.g., shortest paths, centrality, PageRank, and reachability tree) has been studied. However, due to the limited data sources and privacy concerns, proposed approaches are not tested on real-world datasets, but on synthetic datasets by randomly selecting vertices as private ones. Therefore, real-world datasets of public-private networks are essential and urgently needed to such algorithms in the evaluation of efficiency and effectiveness. 

In this paper, we generate four public-private networks from real-world DBLP records, called \ppdblp. We take published articles as public information and regard ongoing collaborations as the hidden information, which are only known by the authors. Our released datasets of \ppdblp offer the prospects for verifying various kinds of efficient public-private analysis algorithms in a fair way. In addition, motivated by widely existing attributed graphs, we propose an advanced model of attributed public-private graphs where vertices have not only private edges but also private attributes. We also discuss open problems on attributed public-private graphs. Preliminary experimental results on our generated real-world datasets verify the effectiveness and efficiency of public-private models and state-of-the-art algorithms. 
\end{abstract}

\section{Introduction}\label{sec.intro}

Online social networks, such as Facebook, Twitter, Google+, Weibo, and Instagram, have been important platforms for the spread of information, ideas, and influence among a huge number of socially connected users. Driven by applications such as social media marketing and user behavior prediction, social network analysis, a process of investigating social structures using network and graph theories, has become a focal point of research. However, privacy issues become a major concern in the algorithmic analysis of social networks. Privacy not only affects the views of a network structure, but also controls the way information shared among social network users. As reported in a recent study \cite{dey2012facebook}, 52.6\% of 1.4 million New York City Facebook users hid their friend's list. Such privacy protection leads to a new graph model, called public-private graphs \cite{chierichetti2015efficient,archer2017indexing,mirzasoleiman2016fast}. It contains a public graph, in which each vertex is also associated with a private graph. The public graph is visible to everyone, and each private graph is visible only to the corresponding user. From each user’s viewpoint, the social network is exactly the union of the public graph and her/his own private graph. Several sketching and sampling approaches \cite{chierichetti2015efficient} have been proposed to address essential problems of graph processing, such as the size of reachability tree \cite{cohen2007summarizing}, all-pair shortest paths \cite{das2010sketch}, pairwise node similarities \cite{haveliwala2002topic}, correlation clustering \cite{bansal2004correlation} and so on.


\begin{figure*}[t]
\vspace{-0.4cm}
\centering
{
\subfigure[Public-private graph]{\includegraphics[width=0.32\linewidth]{\figpath/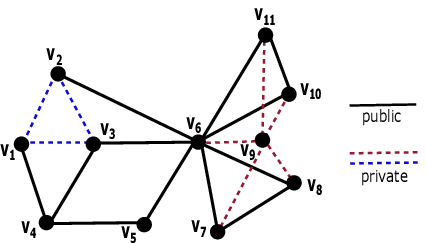}} \hskip 0.15in
\subfigure[Public graph $G$]{\includegraphics[width=0.24\linewidth]{\figpath/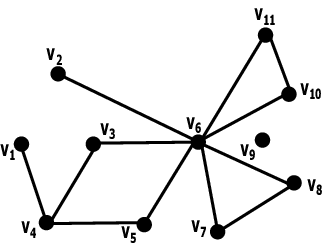} }\hskip 0.1in
\subfigure[In the view of $v_9$: $G\cup G_{v_9}$]{\includegraphics[width=0.24\linewidth]{\figpath/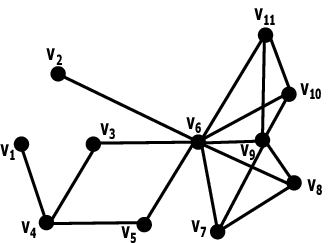} }
}
\caption{An example of undirected and simple public-private graph. In Figure \ref{fig.ppn}(a),  public edges are depicted in solid black lines and private edges are depicted in dashed edges. The red edges incident to vertex $v_9$ are private to $v_9$. The blue edges incident to vertex $v_3$ and the edge $(v_1, v_2)$ are private to $v_3$. Figure \ref{fig.ppn}(b) shows the public graph $G$ consisting of all the solid black edges. Figure \ref{fig.ppn}(c) shows the structure of public-private graph in the view of $v_9$.}
\label{fig.ppn}\vspace{-0.2cm}
\end{figure*}


In the social networks, vertices are always associated with attributes, e.g., a user has information including name, skills, and so on. Recent study \cite{chierichetti2015efficient} focuses on one essential aspect of topological structure of public-private graphs. However, another important issue of vertex attributes has not been investigated yet. In many real-world applications, both the graph topological structure and the vertex properties are important \cite{cheng2012clustering}. In this paper, we model the public-private networks with vertex attributes and give a formulation of attributed public-private networks by considering the public and private vertex attributes.  
More importantly, as far as we know, to date there exist no publicly released datasets of real-world public-private networks. Both \cite{chierichetti2015efficient} and \cite{archer2017indexing} use public graphs to simulate public-private graphs, by randomly selecting vertices and regarding their incident edges as private edges.
Therefore, it is  desirable to have real-world datasets of public-private networks as benchmarks for fair experimental evaluations. This work generates the real-world datasets of public-private networks from real-life DBLP records, denoted by \ppdblp.  We have also publicly released  \ppdblp  to the community.\footnote{\scriptsize{\text{https://github.com/samjjx/pp-data}}} To summarize, this paper makes the following contributions:

\squishlisttight

\item We generate and release a series of real-world public-private network datasets, according to the public and private information on a DBLP network. The released datasets offer the prospects to verify various kinds of public-private graph algorithms in a fair way. We conduct experiments on the \ppdblp datasets to validate the efficiency of state-of-the-art algorithms (Section~\ref{sec.model}).

\item We formally propose a new model of attributed public-private networks, where vertices have private attributes. We highlight two promising directions on the attributed public-private networks and generate the corresponding \ppdblp datasets with attributes from the rich keywords of paper titles on DBLP records (Section~\ref{sec.data}). 
\end{list}

\section{Public-Private Networks}\label{sec.model}

In this section, we first introduce a public-private graph model for online social networks. Then, we generate real-world \ppdblp datasets, and compare state-of-the-art of public-private graph  algorithms on \ppdblp.

\subsection{Public-Private Graph Model}

We present the model of a public-private graph $\mathcal{G}$ \cite{chierichetti2015efficient} as follows. Given a public graph $G=(V,E)$, the vertex set $V$ represents users, and the edge set $E$ represents connections between users. For each vertex $u$ in the public graph $G$, $u$ has an associated private graph $G_u=(V_u, E_u)$, where $V_u \subseteq V$ are the users from public graph and the edge set $E_u$ satisfies $E_u \cap E=\emptyset$. The public graph $G$ is visible to everyone, and the private graph $G_u$ is only visible to user $u$. Thus, in the view of user $u$, she/he can see and access the structure of graph that is the union of public graph $G$ and its own private graph $G_u$ as $G\cup G_u = (V, E\cup E_u)$. Let the private vertex set of $\mathcal{G}$ as $V_{private}=\{u\in V: E_u \neq \emptyset\}$ and the private edge set of $\mathcal{G}$ as $E_{private}=\{(v, w)\in E_u: u\in V_{private}\}$. Note that $E_{private} \cap E=\emptyset$, and each edge presented in different private graphs only counts once in the private edge set $E_{private}$.

\begin{example} 
Consider a pubic-private graph with 11 vertices in Figure \ref{fig.ppn}(a). The pubic-private graph consists of two kinds of edges: public edges and private edges. The public edges are depicted in solid black lines, e.g., $(v_3, v_6)$, indicating that $(v_3, v_6)$ is visible to everyone. The private edges are depicted using dash lines in red and blue.  Edges in red are private and visible to $v_9$, e.g., $(v_6, v_9)$. Edges in blue are private and visible to $v_3$, e.g., $(v_1, v_2)$. Figure \ref{fig.ppn}(b) shows the public graph $G$ consisting of all public edges.  Figure \ref{fig.ppn}(c) shows the structure of  public-private graph in the view of $v_9$, which is richer than the public graph $G$ in Figure \ref{fig.ppn}(b). Because vertex $v_9$  can access all private relationships in private graph $G_{v_9}$. 
\end{example}

\subsection{Constructing  Public-Private DBLP Networks}\label{sub.ppdlbp}
In light of privacy concerns, there exist no publicly released datasets of real-world public-private networks by now. Previous algorithms for public-private social networks, used public graphs to simulate public-private graphs, by randomly selecting some vertices and hiding their incident edges (in a star or a clique) as private edges from the public graph  \cite{chierichetti2015efficient}. To verify competitive algorithms in a fair way, we propose a new approach to generate real-world datasets of public-private DBLP collaboration networks (\ppdblp), according to the public and private information on DBLP records \cite{dblpdataset}.

The intuition is that the information of one accepted paper available in \emph{public} is usually later than the co-author collaboration really happened in \emph{private}. In addition, such collaborations are always only known for authors themselves in person, and are not known to others. Thus, we take collaboration relationships in the published papers as public edges, and regard collaboration relationships in the ongoing works as private edges, which are only known by their authors. Note that if two authors have had a pubic collaboration relationship already, then their private ongoing collaboration is not accounted as  private.

The public-private DBLP network is constructed as follows. We first obtain the DBLP raw data published in 2017 \cite{dblpdataset}. Next, we select one  timestamp $Y$ to distinguish the published yet papers and on-going papers. For example, taking the cut-off timestamp $Y$ as 2013/01/01, all collaborations happened before timestamp $Y$  are regarded as public edges and the collaborations happened on and after timestamp $Y$ are regarded as private edges.  We first construct the public graph. We sort all papers in the increasing order of published dates in DBLP. For each paper $p$ published before timestamp $Y$, we consider each author of this paper as a vertex, and add public edges between any pair of authors in this paper. Similarly, we construct all private graphs as follows. For each paper $p$ published on and after timestamp $Y$, if the two authors $u, v$ do not have a public edge, we add a private edge between $u$ and $v$. This private edge can be accessed by all authors of this paper $p$.
We generate four \ppdblp datasets using four different timestamps $Y$ in \{2013-01-01, 2014-01-01 2015-01-01, 2016-01-01\}. The network statistics of \ppdblp are shown in Table~\ref{tab:dataset}. 

\begin{table}[t]
\vspace*{-0.3cm}
\scriptsize
\caption[]{Network statistics.}\label{tab:dataset}\vspace*{-0.6cm}
\begin{center}%
\begin{tabular}{|l|c|c|c|c|c|}
\hline
{\bf Network} & $|V|$ & $|E|$  & $|V_{private}|$  & $ |E_{private}|$ & $\delta(\mathcal{G})$\\
\hline \hline
\ppdblp-2013 & 1,791,688 & 	5,187,025 &	825,170 &	2,636,570 &	0.086\\  \hline
\ppdblp-2014 & 1,791,688 & 	5,893,083 &	686,292 &	1,930,512 &	0.087\\  \hline
\ppdblp-2015 & 1,791,688 & 	6,605,428 &	515,549 &	1,218,167 &	0.087\\  \hline
\ppdblp-2016 & 1,791,688 & 	7,378,090 &	263,937 &	445,505 &	0.083\\  \hline
\end{tabular}\vspace*{-0.3cm}
\end{center}

\vspace*{-0.1cm}
\end{table}

\subsection{Evaluations on Real-world PP-DBLP datasets}

We use \ppdblp to evaluate two pubic-private graph algorithms proposed by Chierichetti et al. \cite{chierichetti2015efficient}: shortest path approximation and personalized PageRank. Sampling algorithms are developed to precompute the public graph $G$ offline, and then run the online update algorithm on private graph $G_u$ with samples of $G$. We use the publicly available implementation of sampling algorithms\footnote{\scriptsize{\text{https://github.com/aepasto/public-private}}} and set all the same parameters with \cite{chierichetti2015efficient} by default. We randomly select private graphs and report the running time and accuracy of each task averaged over 50 independent tests.  

\begin{figure}[!t]
\centering
{
\subfigure[Efficiency Improvement]{\includegraphics[width=0.47\linewidth]{\figpath/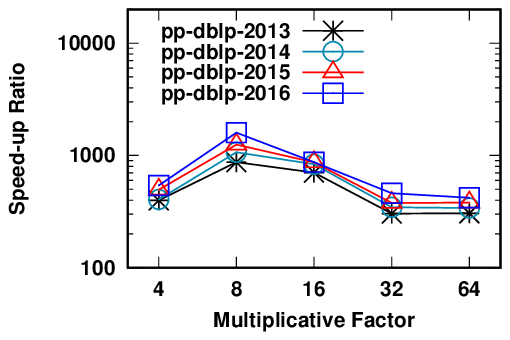}} 
\subfigure[Quality Approximation]{\includegraphics[width=0.47\linewidth]{\figpath/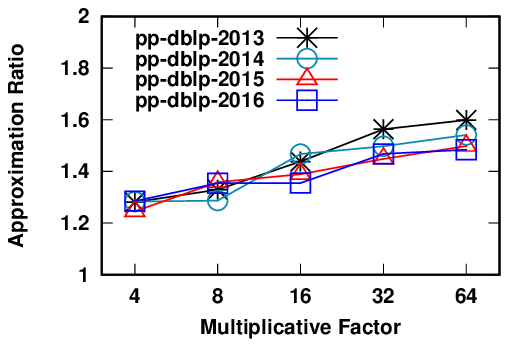} } 
}
\caption{Evaluation of shortest path approximation on \ppdblp networks}
\label{fig.sp}\vspace{-0.2cm}
\end{figure}

\stitle{Shortest Path Approximation.} Given a vertex $u$ in public-private graph $G\cup G_u$, this task is to compute the shortest path from $u$ to another arbitrary vertex in $G\cup G_u$. We evaluate the performance of one sampling algorithm of shortest path \cite{chierichetti2015efficient} with one baseline method of classical Dijkstra's algorithm \cite{cormen2009introduction}.    
Figures \ref{fig.sp}(a) and \ref{fig.sp}(b) respectively show  the results of efficiency improvement and quality approximation varied by the multiplicative factor, which decides the number of samples. The results verify that \cite{chierichetti2015efficient} achieves the great improvement of efficiency (more than 300 times faster than the baseline method) and obtains a good balance of shortest path approximations (no greater than 1.6 times of the optimal answer) on all \ppdblp datasets. Figure \ref{fig.sp}(b) shows that the sampling algorithm \cite{chierichetti2015efficient} improves with smaller approximation ratios as the multiplicative factor decreases, due to the more samples used in the algorithm.

\stitle{Personalized PageRank.} Given a vertex $u$ in public-private graph $G\cup G_u$, the problem of personalized PageRank (PPR) is to find node similarities to $u$ for all vertices in $G\cup G_u$. We compare two methods. The first one is personalized PageRank using heuristic \cite{chierichetti2015efficient}. The second one is a baseline method that  directly applies the algorithm of Andersen et al. \cite{andersen2007local} on graph $G\cup G_u$. The results obtained by the baseline are used as the ground-truth PPR ranking of $u$. Table \ref{tab.pr} reports the efficiency performance and accuracy of personalized PageRank using heuristic on \ppdblp networks. In terms of efficiency comparison, the heuristic algorithm \cite{chierichetti2015efficient} is faster by three orders of magnitude than the baseline \cite{andersen2007local}, which are shown in the column of speed-up ratio in Table \ref{tab.pr}. Table \ref{tab.pr} also shows the accuracy of PPR ranking computed by the heuristic algorithm w.r.t. the ground-truth PPR ranking, in terms of three metrics: the Root Mean Square Error (RMSE), the Cosine Similarity, and the Kendall-$\tau$ index. In terms of RMSE, the heuristic algorithm produces ranking achieving the RMSE close to 0 on all datasets; In terms of cosine similarity, it obtains nearly 1; In terms of the Kendall-$\tau$ correlation of the first 50 positions of the rankings, the score of $\tau@$50 is still quite high falling in [0.537, 0.6067]. Similar results on other datasets are  also reported in \cite{chierichetti2015efficient}.

\begin{table}[t]
\vspace*{-0.2cm}
\scriptsize
\caption[]{Efficiency and accuracy of personalized PageRank using heuristic \cite{chierichetti2015efficient} on \ppdblp networks. }\label{tab.pr}\vspace*{-0.4cm}
\begin{center}%
\begin{tabular}{|l|c|c|c|c|}
\hline
{\bf Network} & Speed-up Ratio & RMSE  & Cosine  & $\tau@$50\\
\hline \hline
\ppdblp-2013 & 6646 & 0.0036 &	0.9907 &	0.6007	 \\ \hline
\ppdblp-2014 & 5746 & 0.0034 &	0.9908 &	0.6067	 \\ \hline
\ppdblp-2015 & 5462 & 0.0041 &	0.9906 &	0.5715	 \\ \hline
\ppdblp-2016 & 6319 & 0.0041 &	0.9890 &	0.5370	 \\ \hline
\end{tabular}\vspace*{-0.2cm}
\end{center}
\vspace*{-0.1cm}
\end{table}

\section{Attributed Public-Private Networks}\label{sec.data}


\begin{figure*}[!t]
\vspace{-0.45cm}
\centering
{
\subfigure[Attributed public-private graph]{\includegraphics[width=0.34\linewidth]{\figpath/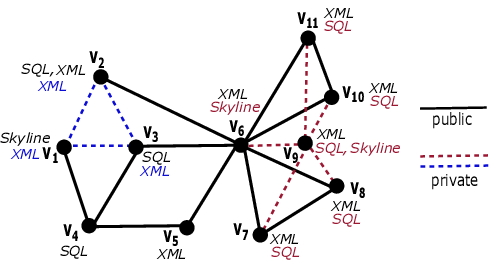}} \hskip 0.15in
\subfigure[Attributed public graph $G$]{\includegraphics[width=0.27\linewidth]{\figpath/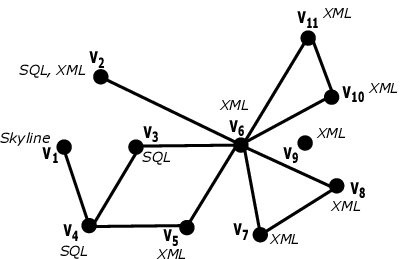} }\hskip 0.1in
\subfigure[In the view of $v_3$: $G\cup G_{v_3}$]{\includegraphics[width=0.27\linewidth]{\figpath/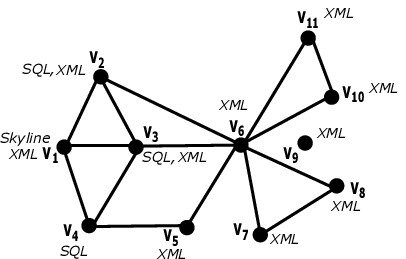} }
}
\caption{An example of attributed public-private graph. In Figure \ref{fig.appn}(a), public attributes are in black. Private attributes are in blue and red, which respectively are visible to $v_3$ and $v_9$.  Figure \ref{fig.appn}(b) shows the attributed public graph $G$ consisting of all public edges and public attributes. Figure \ref{fig.appn}(c) shows the attributed graph $G\cup G_{v_3}$ in the view of $v_3$.}
\label{fig.appn}\vspace{-0.45cm}
\end{figure*}


In this section, we first define an attributed graph model of  public-private networks. We extend the approach of  PP-DBLP generation to produce real-world datasets of attributed public-private networks using title keywords in the papers. Finally, we discuss promising research directions on attributed public-private graphs. 

\subsection{Attributed Public-Private Graph Model}
An attributed public-private graph of $\mathcal{G}$ is modeled as follows. Given an attributed public graph $G=(V,E,A)$, where the vertex set $V$ represents users, the edge set $E$ represents connections between users, and the public attribute set $A(u)$ describes the public attributes of each user $u\in V$. For each user $u$ in the public graph, $u$ has an attributed private graph $G_u=(V_u, E_u, A_u)$, where $V_u \subseteq V$ is a set of users from the public graph, private edge set $E_u \cap E=\emptyset$, and $A_u(v)$ represents the private attributes of vertex $v\in V_u$ that are visible to $u$. The public attributed graph $G$ is visible to everyone, and the private attributed graph $G_u$ is only visible to user $u$. In terms of network structure, attributed public-private graphs have no difference with the public-private graphs in Section \ref{sec.model}. In terms of attributes, consider an attributed public-private graph  $G\cup G_u$, the vertex $u$ can access both public and private attributes of vertex $v$, i.e., $A_v \cup A_u(v)$.

\begin{example}
Figure \ref{fig.appn}(a) shows an example of attributed public-private graph, which has the same graph structure as Figure \ref{fig.ppn}(a). Public attributes are in black, and private attributes are in blue and red. Consider the vertex $v_3$. The public attribute of $v_3$ is $A(v_3)=\{`SQL'\}$. Attributes in blue (e.g., the attribute of `XML' associated with vertices $v_1$, $v_2$ and $v_3$) are private and visible to $v_3$. Thus, $A_{v_3}(v_1)$ $=A_{v_3}(v_2)$ $=A_{v_3}(v_3)$ $=\{'XML'\}$.  Attributes in red (e.g., vertex $v_6$'s attribute of `Skyline') are private and visible to $v_9$. Figure \ref{fig.appn}(b) shows the public attributed graph $G$ consisting of all public edges and public attributes that are visible to everyone. Figure \ref{fig.appn}(c) shows the attributed graph $G\cup G_{v_3}$ in the view of $v_3$. The attributes of $v_1$ are $\{`Skyline', `XML'\}$ as a result of the union of public attributes and private attributes, i.e., $A(v_1) \cup A_{v_3}(v_1)$\xin{, showing that $v_1$ extends her/his  research interests.}
\end{example}

\subsection{Constructing Attributed Public-Private DBLP Networks}
To construct the attributed public-private DBLP networks, we add attributes into vertices on \ppdblp in Section \ref{sub.ppdlbp} as follows. For each author, we collect keywords in the title of all published articles  and extract the most frequent keywords as the public attributes. For the private attributes, let's consider one author $u$ and its attributed private graphs $G_u$. For each author $v$ in $G_u$, the private attributes of $v$ as $A_u(v)$ are the most frequent keywords from the title of all ongoing papers involving authors $v$ and $u$. To select representative keywords, we set each number of public attributes and private attributes is no greater than a maximum threshold of 5, i.e., $|A(v)|\leq 5$ and $|A_u(v)|\leq 5$. The difference of public attributes $A_v$ and private attributes $A_u(v)$ shows the evolving research interests of author $v$. Note that, $A_v \cap A_u(v) \neq \emptyset$ may hold. We use $\theta_u(v)$ to quantify the overlapping ratio of public attributes and private attributes of vertex $v$ in graph $G_{u}$, denoted by $\theta_u(v) =\frac{|A_v \cap A_u(v)|}{|A_v \cup A_u(v)|}$. Let $\delta(u)= \frac{\sum_{v\in V_u} \theta_u(v)}{|V_u|}$ represent the average ratio of overlapping attributes over all authors in $G_u$. For an attributed public-private graph $\mathcal{G}$, we propose $\delta(\mathcal{G})$ to measure the ratio of overlapping public-private attributes  for all private graphs, denoted by $\delta(\mathcal{G})=\frac{\sum_{u\in V_{private}} \delta(u)}{|V_{private}|}$. Table~\ref{tab:dataset} reports the statistic $\delta(\mathcal{G})$ for all \ppdblp datasets.  

\subsection{Open Problems}

We highlight two open problems of keyword search and community search in attributed public-private  networks as follows.

\begin{itemize}

\item  \textbf{Keyword search in attributed public-private networks.}  Keyword search finds users in the vicinity of a given user with similar keywords \cite{zhu2017efficient,techreportICDE}. Keyword search queries in an attributed public-private  network are generated from a vertex that looks for nearest vertices with certain keywords, w.r.t. private structure and private attributes. 

\item  \textbf{Community search in attributed public-private networks.}  
Attributed community search aims at finding the densely-connected subgraphs containing given query nodes with similar attributes \cite{huang2017attribute, huang2017community}. Given a community search query asked by a user $u$, the task needs to be considered in the attributed public-private graph $G\cup G_u$, w.r.t. private structures and private attributes. 
\end{itemize}

\section{Conclusions and Discussions}\label{sec.conclusion}
In this paper, we develop a new model of attributed public-private networks by considering the information of vertices in many real-world networks. In addition, we provide real-world \ppdblp datasets for attributed public-private networks, which are useful to \xin{further research of public-private graphs.}  Besides \ppdblp, one future plan is to construct a real-world pubic-private Facebook social network by conducting a survey of Facebook users, who will be asked to manually identify all of private relationships that they hid. The survey can make use of Facebook built-in app\footnote{\scriptsize{\text{https://apps.facebook.com/my-surveys}}} to conduct investigations.

\section*{Acknowledgment}
This work is supported by the Hong Kong General Research Fund (GRF) Project Nos. HKBU 12200917, 12232716, 12258116, 12632816, and National Natural Science Foundation of China (NSFC) Project Nos. 61702435, 61602395.

\bibliographystyle{IEEEtran}
{\scriptsize
\bibliography{pp-dblp}
}

\end{document}